\documentclass[12pt]{article}%
\usepackage{sw20elba}
\usepackage{amsmath}
\usepackage{graphicx}%
\usepackage{amsfonts}%
\usepackage{amssymb}

\begin{document}

\title{ }

\noindent{\Large What \ High \ Pressure \ Studies \ Have \ Taught \ Us}

\noindent{\Large About \ High-Temperature \ Superconductivity}\textit{\vspace
{0.8cm}}

\noindent JAMES \ S. \ SCHILLING

\noindent\textit{Department of Physics, Washington University}, \textit{C.B.
1105, One Brookings Dr., St. Louis, MO 63130}\vspace{1.2cm}

\noindent\textbf{Abstract\vspace{0.1cm}}

\noindent Superconductivity is an important area of modern research which has
benefited enormously from experiments under high pressure conditions. \ The
focus of this paper will be on three classes of high-temperature
superconductors: \ (1) the new binary compound MgB$_{2},$ (2) the alkali-doped
fullerenes, and (3) the cuprate oxides. \ We will discuss results from
experiment and theory which illustrate the kinds of vital information the
high-pressure variable can give to help better understand these fascinating
materials.\vspace{0.8cm}

\vspace{0.8cm}

\noindent{\Large 1. INTRODUCTION}\textbf{\vspace{0.4cm}}

Superconductivity was discovered in 1911 by G. Holst and Kamerlingh Onnes in
Leiden when elemental Hg was cooled to temperatures below 4.15 K \cite{h1}.
\ Fourteen years later the same group carried out the first high-pressure
experiments on a superconductor \cite{h2}. \ As reported by D. H. Bowen in his
review paper \cite{h3}: ``The first experiments in which stresses were
deliberately applied to superconductors were by Sizoo and Onnes in 1925 who
found that the transition temperatures of tin and indium were increased by
uniaxial tension and decreased by applying pressure to the helium bath in
which they were immersed''. \ In the 76 years since these first experiments,
high-pressure experiments have had a considerable impact on the field of
superconductivity. This is not surprising, since the application of high
pressures has: \ 

\begin{enumerate}
\item led to the discovery of many new superconductors, including 22 elemental
solids\newline (B, O, Si, P, S, Ca, Sc, Fe, Ge, As, Se, Br, Sr, Y, Sb, Te, I,
Cs, Ba, Bi, Ce, Lu) bringing the total number of elemental-solid
superconductors to 51.

\item aided in the synthesis of novel high quality superconducting materials.
This the subject of a paper by C.W. Chu at this conference and will not be
discussed here.

\item guided efforts to enhance the transition temperature $T_{c}$ through
chemical means. \ Even without a detailed understanding of $why\ T_{c}$
changes with pressure, a large magnitude of the pressure derivative
$dT_{c}/dP$ is a good indication that higher values of $T_{c}$ are possible at
ambient pressure through chemical means. \ The observation of a large
enhancement in $T_{c}$ under pressure in the high-temperature oxide
superconductor La-Ba-Cu-O prompted the substitution of the smaller ion
Y$^{3+}$ for La$^{3+}$ to generate lattice pressure, leading to the important
discovery \cite{h6} of superconductivity in YBa$_{2}$Cu$_{3}$O$_{7-\delta}$
(Y-123) at 92 K. \ Below we will see that efforts are being made to expand the
lattice of hole-doped C$_{60}$ in hopes of raising its $T_{c}\simeq$ 52 K to
even higher values \cite{h7}.

\item yielded the dependence of $T_{c}$ on sample volume and lattice
parameters which helped identify the pairing mechanism and test theoretical models.
\end{enumerate}

This paper will focus its attention on this fourth ``benefit'' of high
pressure research. \ The types of superconductor that we consider are the
binary compound MgB$_{2}$, electron- and hole-doped fullerenes, and the
cuprate oxides. \ These materials owe their extraordinary superconducting
properties to their reduced (2D) dimensionality. \ The electron pairing
leading to superconductivity takes place primarily within the B$_{2}$-layers
in MgB$_{2},$ within the CuO$_{2}$-planes in the oxides, and on the surface
(2D!) of the large C$_{60}$ molecule in the doped fullerenes.

The results of high pressure experiments on these important materials can be
best understood if we first consider analogous experiments on conventional
simple metal and transition metal superconductors.\vspace{0.8cm}

\noindent{\Large 2. \ SUPERCONDUCTIVITY \ IN \ CONVENTIONAL \ SUPERCONDUCTORS}%
\textbf{\vspace{0.4cm}}

The fact that high pressure creates 22 new elemental superconductors does
\textit{not} imply that superconductivity is normally enhanced under pressure;
in fact, just the opposite is true. \ The majority of the above
``high-pressure superconductors'' entered this state following a
pressure-induced insulator$\rightarrow$metal transition. \ For most known
superconductors, $T_{c}$ decreases under pressure, sometimes quite rapidly; a
positive value of $dT_{c}/dP$ is rather infrequent. \ The reason for this can
be most easily understood by considering the simple metal superconductors,
like Sn, In, Pb, and Al, where the conduction electrons possess $s,p$
character. \ In all simple metal superconductors, $dT_{c}/dP$ is negative
\cite{r20}: \ for example, Sn (-0.482 K/GPa), In (-0.381 K/GPa), and Pb
(-0.365 K/GPa). \ This ubiquitous decrease in $T_{c}$ is not an electronic
effect, but\ arises predominantly from a stiffening of the lattice with
increasing pressure,\ the same reason the equation-of-state $V(P)$ has an
upward (positive) curvature!

We can make these arguments more specific by considering the BCS expression
for the transition temperature
\begin{equation}
T_{c}\simeq\left\langle \omega\right\rangle \exp\left\{  \frac{-1}%
{N(E_{f})\mathcal{V}}\right\}  ,\vspace{0.1cm}%
\end{equation}
where $\left\langle \omega\right\rangle $ is an average lattice-vibration
frequency, $N(E_{f})$ is the electronic density of states at the Fermi energy,
and $\mathcal{V}$ is the attractive pairing interaction. \ Since the $s,p$
electrons in simple metals are nearly free, one expects approximately
$N(E_{f})\propto V^{+2/3}$ so that $N(E_{f})$ decreases even more slowly than
the sample volume $V$ with increasing pressure. \ However, the principal
reason for the observed decrease in $T_{c}$ with pressure is that the pairing
interaction $\mathcal{V}$ itself decreases by a sizeable amount due to lattice
stiffening, which makes it increasingly difficult for the anions in the
crystal lattice to couple with the electrons. \ 

To put this discussion on a more quantitative basis, we need to consider the
McMillan equation \cite{r17}
\begin{equation}
T_{c}\simeq\frac{\left\langle \omega\right\rangle }{1.20}\exp\left\{
\frac{-1.04(1+\lambda)}{\lambda-\mu^{\ast}(1+0.62\lambda)}\right\}
,\vspace{0.1cm}%
\end{equation}
which connects the value of $T_{c}$ with fundamental parameters such as the
electron-phonon coupling parameter $\lambda,$ an average phonon frequency
$\left\langle \omega\right\rangle ,$ and the Coulomb repulsion which we set
equal to $\mu^{\ast}=0.1$. \ The coupling parameter itself is defined by
$\lambda=N(E_{f})\left\langle I^{2}\right\rangle /[M\left\langle \omega
^{2}\right\rangle ],$ where $\left\langle I^{2}\right\rangle $ is the average
squared electronic matrix element, $M$ the molecular mass, and $\left\langle
\omega^{2}\right\rangle $ the average squared phonon frequency. \ Taking the
logarithmic volume derivative of $T_{c}$ in Eq. (2), we obtain the simple
relation
\begin{equation}
\frac{d\ln T_{c}}{d\ln V}=-B\frac{d\ln T_{c}}{dP}=-\gamma+\Delta\left\{
\frac{d\ln\eta}{d\ln V}+2\gamma\right\}  ,
\end{equation}
where $B$ is the bulk modulus, $\gamma\equiv-d\ln\left\langle \omega
\right\rangle /d\ln V$ the Gr\"{u}neisen parameter, $\eta\equiv N(E_{f}%
)\left\langle I^{2}\right\rangle $ \cite{r17''}, and $\Delta\equiv
1.04\lambda\lbrack1+0.38\mu^{\ast}]\left[  \lambda-\mu^{\ast}(1+0.62\lambda
)\right]  ^{-2}$. \ Eq. (3) has a simple interpretation. \ The first term on
the right, which comes from the prefactor to the exponent in the above
McMillan expression for $T_{c}$, is usually small relative to the second term.
\ The sign of the pressure derivative $dT_{c}/dP$, therefore, is determined by
the relative magnitude of the two terms in the curly brackets.

The first ``electronic'' term in the curly brackets involves the derivative of
the Hopfield parameter $\eta\equiv N(E_{f})\left\langle I^{2}\right\rangle $,
an ``atomic'' property which can be calculated directly in band-structure
theory. \ In his landmark paper \cite{r17}, McMillan demonstrated that whereas
individually $N(E_{f})$ and $\left\langle I^{2}\right\rangle $ may fluctuate
appreciably, their product $\eta\equiv N(E_{f})\left\langle I^{2}\right\rangle
$ changes only gradually, i.e. $\eta$ is a well behaved ``atomic'' property.
\ One would thus anticipate that $\eta$ changes in a relatively well defined
manner under pressure, reflecting the character of the electrons near the
Fermi energy. \ An examination of the body of high-pressure data on simple
$s,p$-metal superconductors, in fact, reveals that $\eta$ normally increases
under pressure at a rate close to $d\ln\eta/d\ln V\approx-1$ \cite{r17'}.
\ For transition-metal (d-electron) superconductors, on the other hand,
Hopfield has pointed out \cite{r17''} that the larger value $d\ln\eta/d\ln
V\approx-3$ to -4 is more appropriate.

Let us now apply Eq. (3) to an analysis of $dT_{c}/dP$ for simple metal
superconductors. \ The second ``lattice'' term in the curly brackets in Eq.
(3) is positive since the lattice term is positive ($2\gamma\approx+3$ to
$+5)$ and dominates over the negative electronic term $d\ln\eta/d\ln
V\approx-1$. \ Since\ $\Delta$ is always positive and -$\gamma$ is relatively
small, the sign of $dT_{c}/dP$ is negative, opposite to that in the curly
brackets. \ This accounts for the universal decrease of $T_{c}$ with pressure
due to lattice stiffening in simple metals. \ In Sn, for example, $T_{c}$
decreases under pressure at the rate $dT_{c}/dP\simeq$ -0.482 K/GPa which
leads to $d\ln T_{c}/d\ln V\simeq+7.2$ \cite{r20}. \ Inserting for Sn
$T_{c0}\simeq$ 3.73 K, $\left\langle \omega\right\rangle \simeq110$ K
\cite{r21}, and $\mu^{\ast}=0.1$ into the above McMillan equation, we obtain
$\lambda\simeq0.69$ from which follows that $\Delta\simeq2.47.$ \ Inserting
the above values into Eq. (3) and setting $d\ln\eta/d\ln V\approx-1$, we can
solve Eq. (3) for the Gr\"{u}neisen parameter to obtain $\gamma\simeq+2.46,$
in reasonable agreement with the experimental value $\gamma\approx+2.1$
\cite{r20}. \ Similar results are obtained for other conventional simple metal
BCS superconductors.

In transition metal superconductors the electrons taking part in the
superconductivity have predominantly $d$ character which often leads to higher
values of the density of states $N(E_{f})$ and transition temperature $T_{c}.$
\ In many transition metals $T_{c}$ decreases with pressure, but in some
$T_{c}$ increases. \ Indeed, under pressure $N(E_{f})$ can either decrease or
increase; should $E_{f}$ lie on the low energy side of a peak in $N(E),$
$s\rightarrow d$ electron transfer under pressure would lead to an increase in
$N(E_{f}),$ and vice versa should $E_{f}$ lie on the high energy side of a
peak. \ The moderating influence of the change in $\left\langle I^{2}%
\right\rangle $ under pressure leads to the universal increase of their
product $\eta\equiv N(E_{f})\left\langle I^{2}\right\rangle $ according to
$d\ln\eta/d\ln V\approx-3$ to -4, as pointed out by Hopfield \cite{r17''}.
\ If this relatively large electronic term becomes larger than the lattice
term 2$\gamma$ in Eq. (3), $T_{c}$ would be expected to \textit{increase} with
pressure; this is, in fact, observed in the transition metals V \cite{smith1}
and La \cite{smith2}, among others. \ Unlike for $s,p$ metals, the pressure
dependence $T_{c}(P)$ for transition metals follows no universal behavior,
reflecting the additional complexity, and potency, of the electronic
properties in a $d$ electron system.\vspace{0.8cm}

\noindent{\Large 3. \ SUPERCONDUCTIVITY \ IN MgB}$_{2}$\textbf{\vspace{0.4cm}}

\noindent The discovery of superconductivity at the high temperature
$T_{c}\approx40$ K in the simple $s,p$-metal compound MgB$_{2}$ was quite
unexpected \cite{n4}. \ The absence \cite{n5} of the problematic weak-link
behavior of the high-$T_{c}$ oxides and the relative ease of synthesis in
various forms \cite{n6} has raised hopes that MgB$_{2}$ may be suitable for
numerous technological applications. \ Efforts to enhance the value of $T_{c}
$ in this class of superconductor would be aided by the identification of the
superconducting mechanism and by establishing systematics in the
superconducting and normal-state properties. \ MgB$_{2}$ is a quasi-2D
material with strong covalent bonding within the boron layers. \ It is thus
not surprising that the compression is anisotropic \cite{r9,r8,r10,goncharov},
the most accurate structural measurements \cite{r10} revealing that under
hydrostatic pressure the initial compression along the $c$ axis is 64\%
greater than along the $a$ axis; the bulk modulus is $B=147.2(7)$. \ The
anisotropy in the superconducting properties is also appreciable, the upper
critical field ratio $H_{c2}^{ab}/H_{c2}^{c}$ reportedly being $2-3$
\cite{n12,lee}, less than that observed in the high-$T_{c}$ oxides
\cite{n12'}.\ \ A full characterization of the remaining anisotropic
properties awaits the synthesis of sufficiently large single crystals.

Soon after the discovery of superconductivity in MgB$_{2},$ three groups
\cite{n14,n14',monte} reported that $T_{c}$ decreased under high pressure, but
the rate of decrease varied considerably from -1.6 K/GPa \cite{n14} to -1.9
K/GPa \cite{n14'} in piston-cylinder studies with Fluorinert pressure medium,
to $\sim$ -0.6 K/GPa in quasi-hydrostatic studies \cite{monte} to 25 GPa with
solid steatite pressure medium. \ Choi \textit{et al}. \cite{choi} have
recently carried out resistivity studies to 1.5 GPa pressure in
daphne-kerosene pressure medium, obtaining $dT_{c}/dP\simeq-1.36$ K/GPa. \ The
differing pressure dependences may be due to differences in the samples and/or
to shear stress effects in the frozen or solid pressure media.

We recently carried out He-gas hydrostatic pressure experiments \cite{n18} on
the same high quality isotopically pure ($^{11}$B) sample used in the above
structural studies \cite{r10}. \ For $P\lesssim0.5$ GPa helium is fluid at
$T_{c}\approx39$ K. \ At higher pressures the shear stresses are held to a
minimum by the carefully controlled manner \cite{schirber1} in which solid
helium is allowed to freeze around the sample. \ The dependence of $T_{c}$ on
hydrostatic pressure is seen in Fig. 1(a) to be highly linear, $dT_{c}%
/dP\simeq-1.11(2)$ K/GPa (yielding $d\ln T_{c}/d\ln V=Bd\ln T_{c}%
/dP\simeq+4.16),$ and does not depend on the pressure/temperature history of
the sample. \ Such history effects are rare in superconductors without
pressure-induced phase transitions, but do occur in certain high-$T_{c}$
oxides containing defects with appreciable mobility at RT \cite{relaxation}.
\ In addition, we observed that $dT_{c}/dP$ remained unchanged if neon was
substituted for He as pressure medium, confirming the absence of intercalation
effects in MgB$_{2}$. \ Lorenz \textit{et al.} \cite{lorenz2} have very
recently carried out He-gas studies to 0.8 GPa on two MgB$_{2}$ samples with
differing $T_{c0}$ values 39.2 K and 37.5 K obtaining $dT_{c}/dP\simeq-1.07$
K/GPa and $-1.45$ K/GPa, respectively. \ These authors conclude that
differences in the samples themselves, and not shear stress effects, are
responsible for the differing $dT_{c}/dP$ values.

In Fig. 1(b) we show the dependence of $T_{c}$ on pressure to 20 GPa for
MgB$_{2}$ using a diamond-anvil-cell (DAC) with dense helium pressure medium
\cite{n19}, thus extending the pressure range of the above He-gas studies
nearly thirtyfold. \ $T_{c}$ is seen to decrease approximately linearly with
pressure to 10 GPa, consistent with the rate -1.11 K/GPa (dashed line). \ Very
recently Tissen \textit{et al.} \cite{tissen3} have carried out ac
susceptibility measurements in a DAC to 28 GPa on a MgB$_{2}$ sample with
$T_{c0}\simeq$ 37.3 K at ambient pressure. \ They find the high initial slope
$dT_{c}/dP\simeq-2$ K/GPa, $T_{c}$ decreasing to 11 K at 20 GPa and 6 K at 28
GPa. \ They also report that the pressure dependence $T_{c}(P)$ shows a bump
near 9 GPa which they speculate may arise from an electronic Lifshitz
transition. \ Further experiments are necessary to determine whether this bump
is intrinsic to the sample or the result of shear stress effects in the frozen
methanol-ethanol pressure medium.%

\begin{center}
\includegraphics[
trim=0.179907in 0.180292in 0.000000in 0.121231in,
natheight=7.771200in,
natwidth=9.468800in,
height=2.4249in,
width=3.0096in
]%
{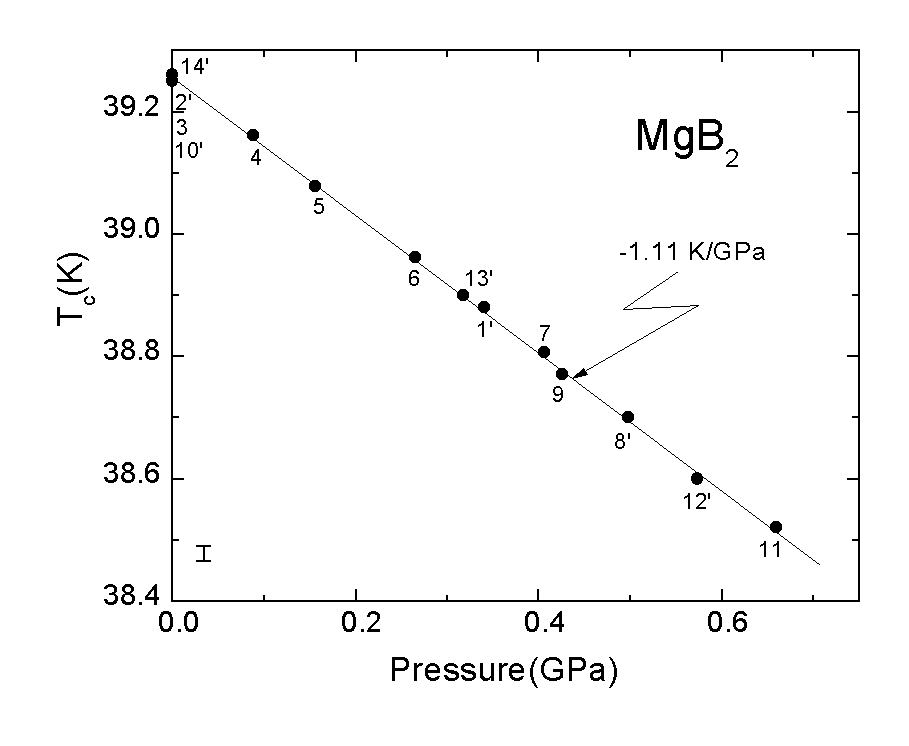}%
\end{center}

\begin{center}
\includegraphics[
natheight=7.937300in,
natwidth=9.416900in,
height=2.5728in,
width=3.0476in
]%
{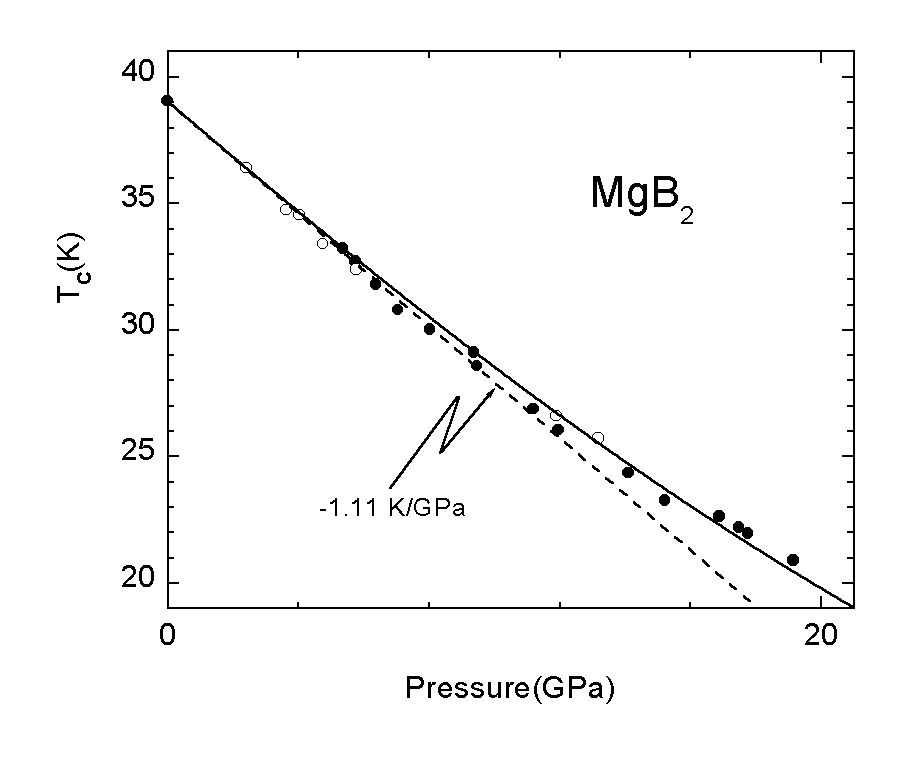}%
\end{center}

\noindent{\small Fig. 1. \ (a)(top) Superconducting transition temperature
onset versus applied pressure from Ref. \cite{n18}. \ Numbers give order of
measurement. \ A typical error bar for }$T_{c}${\small \ (}$\pm0.01$%
{\small \ K) is given in lower left corner; the error in pressure is less than
the symbol size. \ Pressure was either changed at RT (unprimed numbers) or at
low temperatures }$\sim${\small \ 60 K (primed numbers). (b)(bottom)
\ Superconducting transition temperature midpoint versus pressure to 20 GPa
from diamond-anvil-cell measurements in Ref. \cite{n19}. \ Data taken first
for monotonically increasing pressure (}$\bullet),${\small \ then for
monotonically decreasing pressure (}$\circ${\small ). \ The straight dashed
line has slope -1.11 K/GPa.}\vspace{0.6cm}

We now apply the same analysis carried out above for simple metal
superconductors to MgB$_{2}$ to see whether $T_{c}(P)$ from our measurements
is consistent or not with BCS electron-phonon coupling theory. \ We first
analyze the He-gas data which gives the initial dependence. \ Using the
average phonon energy from inelastic neutron studies \cite{n9} $\left\langle
\omega\right\rangle =670$ K, $T_{c0}\simeq39.25$ K, and $\mu^{\ast}=0.1,$ we
obtain from the above relations $\lambda\simeq0.90$ and $\Delta\simeq1.75$.
\ Our estimate of $\lambda\simeq0.90$ agrees well with those of other authors
\cite{kong,an}. \ Since the pairing electrons in MgB$_{2}$ are believed to be
$s,p$ in character \cite{kortus,medvedeva,kong,neaton}, we set $d\ln\eta/d\ln
V\approx-1,$ a value close to $d\ln\eta/d\ln V=Bd\ln\eta/dP\approx-0.81,$
where $B=147.2$ GPa from Ref. \cite{r10} and $d\ln\eta/dP\approx+0.55$ \%/GPa
from first-principles electronic structure calculations by Medvedera
\textit{et al.} \cite{note10}. \ Inserting the above values of $d\ln
T_{c}/d\ln V=+4.16$, $\Delta=1.75$, and $d\ln\eta/d\ln V=-1$ into Eq. (3), we
find $\gamma=2.36,$ in reasonable agreement with the value $\gamma\approx2.9$
from Raman spectroscopy studies \cite{goncharov} or $\gamma\approx2.3$ from
\textit{ab initio }electronic structure calculations on MgB$_{2}$ \cite{r18}.
\ The He-gas $T_{c}(P)$ data are thus clearly consistent with electron-phonon
pairing in MgB$_{2}.$

A more stringent test of this conclusion is provided by the DAC $T_{c}(P)$
data in Fig. 1(b) which cover a relatively large $\sim$ 10\% change in volume.
\ We would like to see whether we can reproduce the DAC data using the
McMillan equation and suitably extrapolating the parameters used in the above
analysis of the He-gas data. \ As pointed out by Chen \textit{et al.}
\cite{chen2}, an appropriate method of extrapolation is to integrate the
volume derivatives of the above parameters $\gamma\equiv-d\ln\left\langle
\omega\right\rangle /d\ln V=$ +2.36 and $d\ln\lambda/d\ln V=d\ln\eta/d\ln
V-d\ln\left\langle \omega^{2}\right\rangle /d\ln V=-1-2(-2.36)=+3.72$ to
obtain $\left\langle \omega\right\rangle =($670 K$)(V/V_{0})^{-2.36}$ and
$\lambda=0.90(V/V_{0})^{3.72}.$ \ Inserting these two volume dependences in
the McMillan equation, and assuming $\mu^{\ast}=0.1$ is independent of
pressure \cite{chen2}, we obtain the dependence of $T_{c}$ on relative volume
$V/V_{0}$. \ This can be converted to the dependence of $T_{c}$ on pressure
$P$ by using\ the Murnaghan equation-of-state $V(P)/V_{0}=[1+B^{\prime
}P/B]^{-1/B^{\prime}}$ where we use the value $B=147.2$ GPa from Ref.
\cite{r10} and the canonical value $B^{\prime}\equiv dB/dP=4$ supported by a
recent calculation \cite{loa}. \ As seen in Fig. 1(b), the agreement of this
calculated $T_{c}(P)$ dependence (solid line) with the experimental data is
quite impressive. \ According to this estimate, a pressure of $P\approx50$ GPa
would be required to drive $T_{c}$ to below 4 K. \ A similar calculation was
very recently carried out by Chen \textit{et al.} \cite{chen2} over a much
wider pressure range; this paper also contains a detailed discussion of the
pressures dependences of $\eta,$ $\lambda,$ and $\mu^{\ast}.$

The good agreement between the experimental data to 20 GPa and the predictions
of the McMillan formula using the volume dependences determined from the
He-gas high-pressure data to 0.7 GPa provides substantial evidence that
superconductivity in MgB$_{2}$ originates from standard BCS phonon-mediated
electron pairing. \ This finding agrees with high precision\ isotope effect
experiments \cite{n7,n7'}, among others. \ The fact that the B isotope effect
is fifteen times that for Mg \cite{n7'} is clear evidence that the
superconducting pairing originates within the graphite-like B$_{2}%
$-layers\vspace{0.8cm}

\noindent{\Large 4. \ SUPERCONDUCTIVITY \ IN \ THE \ ALKALI-DOPED
\ FULLERENES}\textbf{\vspace{0.4cm}}

\noindent Another class of superconductors having very high values of $T_{c}$
are the alkali-doped fullerides A$_{3}$C$_{60}$, where A = K, Rb, Cs, or some
combination thereof \cite{f87}. \ As for the simple metal and MgB$_{2}$
superconductors, $T_{c}$ for the alkali-doped fullerides is found to decrease
under the application of hydrostatic pressure \cite{f1}. \ For example, for
Rb$_{3}$C$_{60}$, where $T_{c0}\simeq29.5$ K, we obtained $dT_{c}%
/dP\simeq-8.7$ K/GPa \cite{f2}. \ From our measurement of the bulk modulus in
a neutron diffraction experiment \cite{f3}, $B=18.3$ GPa, we estimated $d\ln
T_{c}/d\ln V=Bd\ln T_{c}/dP\simeq+5.4,$ a value intermediate between that for
MgB$_{2}$ (+4.16) and that for Sn (+7.2). \ One is tempted to account for the
decrease in $T_{c}$ with pressure for Rb$_{3}$C$_{60}$ or the other A$_{3}%
$C$_{60}$ fullerides by invoking pressure-induced lattice stiffening. \ Such
an attempt, however, fails. \ In contrast to MgB$_{2}$ and Sn, the electronic
bandwidth is quite narrow, leading to the expectation of a sizeable decrease
in $N(E_{f})$ under pressure. \ To determine the reason for the large negative
value of $dT_{c}/dP$, we measured in a single experiment the pressure
dependence of both $T_{c}$ and $N(E_{f})$ for Rb$_{3}$C$_{60},$ determining in
the process the functional dependence of $T_{c}$ on $N(E_{f})$ \cite{f2}. \ A
detailed analysis revealed that weak-coupling theory can account for these
pressure dependences as long as the characteristic energy of the intermediary
boson is between 300 K and 800 K, typical energies for the high frequency
on-ball phonons. \ The mechanism behind the large negative value of
$dT_{c}/dP$ in Rb$_{3}$C$_{60}$ is thus not lattice stiffening, as in the
simple metals and MgB$_{2},$ but a sharp decrease in the electronic density of
states $N(E_{f}) $ with pressure.

This result sheds light on the observed increase in $T_{c}$ in A$_{3}$C$_{60}$
as the larger Rb$^{+1}$ ion is substituted for the smaller K$^{+1}$, thus
expanding the lattice, as seen in Fig. 2. In fact, it was widely believed that
the relationship between $T_{c}$ and the lattice parameter $a$ followed a
universal behavior in both cation substitution and high-pressure experiments.
\ Our combined equation-of-state and $T_{c}(P)$ studies on Rb$_{3}$C$_{60}$,
however, revealed that two different dependences are found, as seen in Fig. 2.
\ The reason for this effect is not known, but may have to do with different
rotationally ordered states of the C$_{60}$ molecule in the two cases which
influences the density of states and thus $T_{c}$.%

\begin{center}
\includegraphics[
natheight=20.823000in,
natwidth=29.958000in,
height=1.9977in,
width=2.8677in
]%
{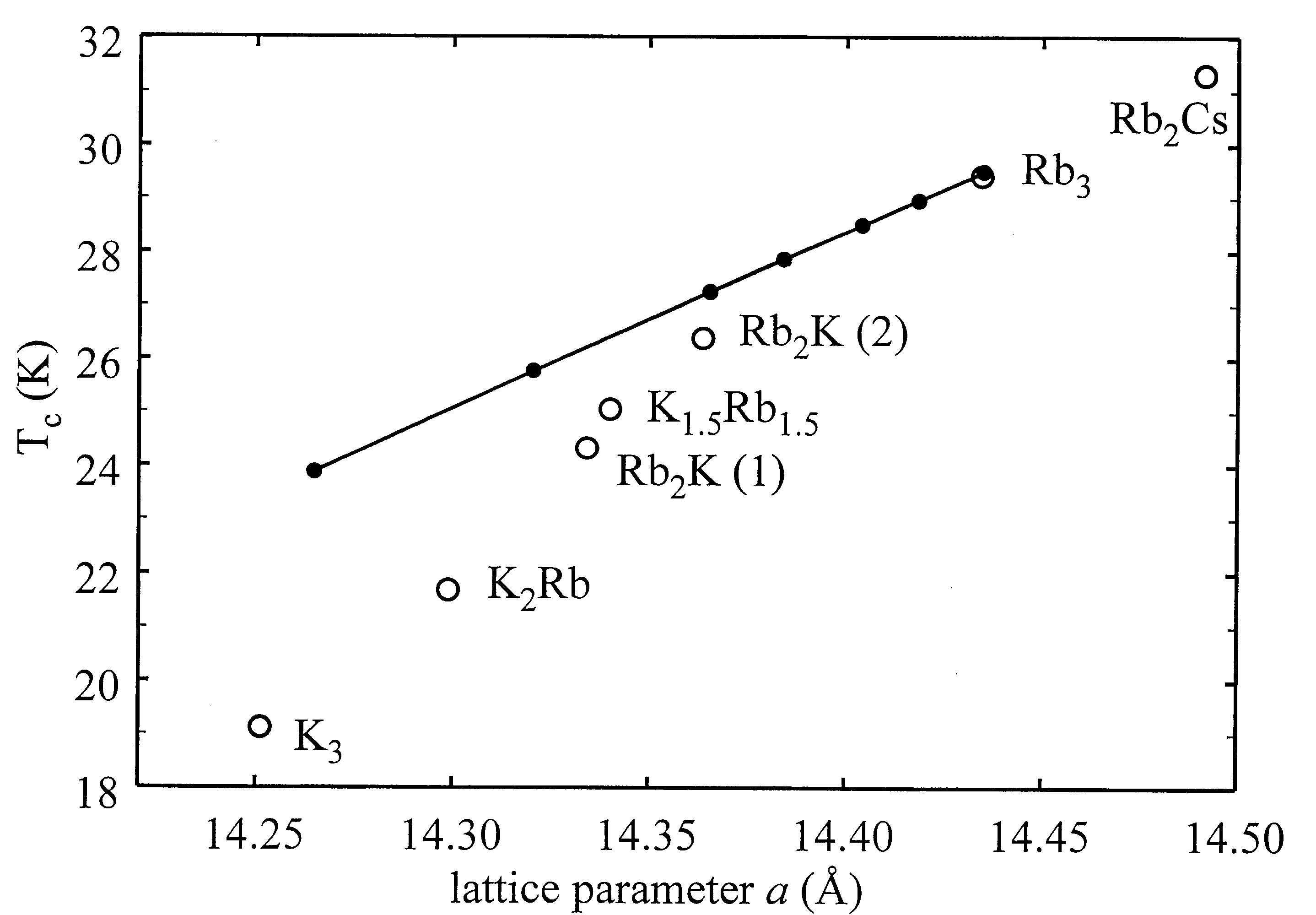}%
\end{center}

\noindent{\small Fig. 2. \ T}$_{c}$ {\small versus lattice parameter from Ref.
\cite{f3} for both (}$\bullet${\small ) high-pressure experiments on Rb}$_{3}%
${\small C}$_{60}${\small \ and (}$\circ${\small ) cation-substitution
experiments. \ Only the cation(s) to C}$_{60}${\small \ are listed in the
figure. \ Solid line is guide to eye.}\vspace{0.6cm}

Using gate-induced doping in a field-effect transistor configuration,
Sch\"{o}n \textit{et al}.\ have demonstrated superconductivity in both
electron- \cite{schoen} and hole-doped \cite{schoen2} C$_{60}$ single crystals
with lattice parameter 14.16 \AA. \ The maximum values of $T_{c}$ are found to
be 11 K and 52 K, respectively. \ The latter value is a record for a
non-cuprate superconductor. \ The higher value of $T_{c}$ for hole doping
compared to electron doping is apparently correlated with the higher density
of states in the valence band (HOMO) compared with the conduction band (LUMO).

From the above discussion, $T_{c}\simeq52$ K for optimally hole-doped C$_{60}$
would evidently be expected to increase further if its lattice parameter $a$
could be expanded. \ We found above for Rb$_{3}$C$_{60}$ that $d\ln T_{c}/d\ln
V=d\ln T_{c}/3d\ln a\simeq+5.4$, which implies that $d\ln T_{c}/d\ln
a\simeq+16.2$. \ This means that one would expect $T_{c}$ to increase about 16
times faster than the lattice parameter $a.$ \ If we expand the lattice
parameter $a$ of C$_{60}$ by 2\%, so that it equals that of Rb$_{3}$C$_{60}%
,$\ one would expect $T_{c}$ to increase by approximately $16.2\times
2\%\simeq32\%,$ implying an increase in $T_{c}$ from 52 K to nearly 70 K. \ To
test the above, it would be of interest to expand C$_{60}$'s lattice, perhaps
through suitable intercalation of rare gas atoms.\vspace{0.8cm}

\noindent{\Large 5. \ SUPERCONDUCTIVITY \ IN \ THE \ CUPRATE \ OXIDES}%
\textbf{\vspace{0.4cm}}

\noindent It has been nearly 15 years since the discovery \cite{n1} of
high-$T_{c}$ superconductivity in the cuprate oxide La-Ba-Cu-O at
$T_{c}\approx35$ K. \ In spite of enormous efforts since then, including over
50,000 experimental and 20,000 theoretical papers, there is still no consensus
on the underlying mechanism responsible for the superconducting pairing.
\ This is one of the great unsolved problems in Condensed Matter Physics.
\ The maximum value of $T_{c}$ has improved from initially $\sim$ 35 K for
La-Ba-Cu-O and related systems to 134 K for HgBa$_{2}$Ca$_{2}$Cu$_{3}%
$O$_{8+\delta}$ (Hg-1223) at ambient pressure in 1993 \cite{hg} and $\sim$ 160
K for the same compound at 30 GPa pressure in 1994 \cite{gao1}. \ There have
even been unconfirmed reports of superconductivity at 250 K in Tl$_{2}$%
Ba$_{2}$Ca$_{2}$Cu$_{3}$O$_{10+\delta}$ (Tl-2223) above 4 GPa \cite{han} and
at 330 K in the Pb-Ag-C-O system at ambient pressure \cite{djurek}.

Unfortunately, Nature has been particularly inventive in her efforts to thwart
understanding. \ The distorted modified perovskite structure of these
materials undergoes subtle structural changes when temperature or pressure is
varied which complicates the interpretation of experimental data. \ Defects of
many different kinds have a profound influence on the superconducting
properties, including the value of $T_{c}.$ \ An example is the emergence of
the low-temperature-tetragonal (LTT) phase in La$_{2-x}$Ba$_{x}$CuO$_{4}$ for
$x\approx0.125$ which completely suppresses $T_{c}$ from its nominal value of
$\sim37$ K. \ Because of the strong inverse parabolic dependence of $T_{c}$ on
the carrier concentration $n$ seen in Fig. 3 \cite{presland1}, small changes
in $n$ due to the influence of defects can have a surprising large effect on
$T_{c}.$

In spite of the complexity of these materials, researchers \cite{jorg1} have
been able to identify a number of important guidelines for enhancing the value
of $T_{c}$ in the superconducting oxides: \ (1) the carrier concentration $n$
in the CuO$_{2}$ planes should be varied through suitable cation substitution
until its optimal value is reached (see Fig. 3); (2) try to increase the
number of CuO$_{2}$ planes which lie close together (in a packet) in the oxide
structure while maintaining optimal doping - \ ``respectable'' one-plane
systems, like Tl-2201, have $T_{c}$ values in the range 90 - 100 K, two-plane
systems in the range 100 - 120 K, and three-plane systems in the range 120 -
140 K; (3) try to position defects as far from the CuO$_{2}$ planes as
possible; and (4) since $T_{c}$ is diminished with increasing buckling angle
in the CuO$_{2}$ planes, develop structures where the CuO$_{2}$ planes are as
flat as possible.

We would like to pose the following question: \ are there perhaps other
guidelines to maximize $T_{c}$ which high pressure studies can give us? \ Can
high pressure experiments give us NEW information not available from other
studies? \ The most studied superconducting property under pressure
is\ $T_{c}(P)$ which depends on the system studied, the doping level $n,$ the
type and mobility of defects, and, in some cases, on the pressure medium used
\cite{r17',d2}. \ However, considering the totality of $T_{c}(P)$ data, one
feature stands out: \ that more often than not $T_{c}$ \textit{increases} with
pressure, as first recognized by Schirber \cite{schirber4}. \ The foregoing
discussion should help the reader appreciate that an increase in $T_{c}$ with
pressure is something special! \ In the case of the optimally doped Hg
compounds with one, two and three CuO$_{2}$ planes, the initial rate of
increase is identical, $dT_{c}/dP\approx+1.75$ K/GPa \cite{klehe1}. \ It is
this constant but perservering increase in $T_{c}$ with pressure which allows
$T_{c}$ in Hg-1223 to increase from 134 K at ambient pressure to $\sim$ 160 K
at 30 GPa at which pressure $T_{c}(P)$ passes through a maximum \cite{gao1}.

In the cuprate oxides, the canonical change in $T_{c}$ with pressure is that
it first increases with pressure, passes through a maximum value at some critical%

\begin{center}
\includegraphics[
natheight=17.781401in,
natwidth=24.198299in,
height=2.6308in,
width=3.5743in
]%
{figures/Tc(n)3.png}%
\end{center}

\noindent{\small Fig. 3. \ Inverted parabolic dependence of T}$_{c}$
{\small on hole-carrier content for the superconducting oxides adapted from
Ref. \cite{presland1}. \ Representative experimental values of dT}$_{c}%
${\small /dP for underdoped, optimally doped and overdoped sampes are
given.\vspace{0.6cm}}

\noindent pressure, and then decreases \cite{r17',d2}. \ What causes this
$T_{c}(P)$ dependence? \ An examination of Fig. 3 leads one to the idea that
perhaps the hole-carrier concentration $n$ increases with pressure so that for
an underdoped sample, where $n<n_{opt},$ $T_{c}(P)$ simply tracks the
canonical bell-shaped $T_{c}(n)$ curve. \ Indeed, Hall effect studies show
that $n$ normally increases with pressure at the rate +10\%/GPa \cite{r17'}.
\ If this were the whole story, then for an optimally doped sample one would
expect $dT_{c}/dP=(dT_{c}/dn)(dn/dP)=0,$ since $T_{c}(n)$ is at an extremum
for $n=n_{opt}$ (see Fig. 3). \ As indicated in Fig. 3, however, this
expectation from this simple charge-transfer model is not confirmed in
experiment. \ For ``normal'' high-$T_{c}$ oxides, $dT_{c}/dP\approx+1$ to $+2$
at optimal doping! \ This means that there are at least two effects
determining the total pressure dependence of $T_{c}$%
\begin{equation}
\frac{dT_{c}}{dP}=\left(  \frac{dT_{c}}{dP}\right)  ^{intrinsic}+\left(
\frac{dT_{c}}{dn}\right)  \left(  \frac{dn}{dP}\right)  ,
\end{equation}
an intrinsic dependence ($dT_{c}/dP)^{intrinsic}\approx+1$ to $+2,$ and a
second dependence arising from pressure-induced changes in the carrier
concentration $n$ in the CuO$_{2}$ planes. \ It is this intrinsic term which
promises to tell us something new about high-$T_{c}$ superconductors,
something we perhaps could not have learned from ambient pressure experiments
under varying sample stoichiometries and structure types. \ However, we have
only evaluated ($dT_{c}/dP)^{intrinsic}$ at optimal doping. \ It is certainly
possible that this intrinsic dependence varies with the carrier concentration.
\ To extract the intrinsic term from $T_{c}(P)$ data with varying values of
$n,$ we must be able to estimate the second term on the right in Eq. (4), the
charge transfer term. \ This is no mean feat since this term depends on both
the rate of change of $n$ with pressure, as well as $dT_{c}/dn.$

There are, however, further problems which make the extraction of the
intrinsic term even more difficult. \ In many superconducting oxides the
application of pressure doesn't simply compress the lattice, but also prompts
mobile oxygen defects to assume a greater degree of local order
\cite{r17',klehe2}. \ This leads to relaxation effects which are both
temperature and pressure dependent. \ The oxygen chain sublattice in YBa$_{2}%
$Cu$_{3}$O$_{7-y}$ (Y-123), for example, is partially occupied with oxygen
anions with a considerable mobility at room temperature. \ This allows these
oxygen defects to assume a myriad of different substructures or local ordered
states which can influence the value of $T_{c}$. \ Under pressure the oxygen
defects may migrate from one local ordered state to another, but at a
progressively slower rate, the higher the pressure \cite{sade1}. \ In fact,
one can measure the pressure dependence of the relaxation time $\tau(P)$ and
use this to estimate the most likely diffusion path of oxygen defects through
the solid \cite{sade3}. \ For studies of the $T_{c}(P)^{intrinsic},$ however,
these relaxation effects are more than just a nuisance. \ One strategy to
eliminate the relaxation effects completely is to carry out the entire high
pressure experiment at low temperatures, allowing $\tau,$ which depends
exponentially on temperature, to become extremely large. \ The hope would be
that the elimination of relaxation effects would turn the complex dependency
$T_{c}(n,P)$ into a simple one. \ Unforunately, experiments to 20 GPa in a DAC
on Y-123 carried out solely at temperatures low enough (%
$<$
90 K) to suppress oxygen ordering brought little simplification \cite{sade1}.
\ In the Y-123 system, at least, the extraction of the dependence
$T_{c}(P)^{intrinsic}$ would seem out of reach.

Fortunately, these relaxation effects operate by changing the carrier
concentration $n.$ \ By considering pressure derivatives $dT_{c}/dP$ only for
optimally doped samples, we can eliminate not only any changes in $T_{c}$ from
the normal increase in $n$ with pressure, but also those changes arising from
oxygen ordering phenomena. \ Perhaps the most important fundamental result
from all high pressure experiments on the optimally doped oxides is that
$dT_{c}/dP=(dT_{c}/dP)^{intrinsic}\approx+1.5$ K/GPa which corresponds to the
volume dependence $T_{c}\propto V^{-1.2}$ or \newline $d\ln T_{c}/d\ln
V\simeq-1.2$ \cite{r17'}$,$ a far weaker volume dependence than obtained for
the $s,p$ metal systems Sn (+7.2) or MgB$_{2}$ (+4.16) where $T_{c}$ increases
with pressure. \ This comparison emphasizes how small the percent change in
$T_{c}$ under pressure really is for the superconducting oxides. \ The fact
that numerous different systems yield approximately the same volume dependence
speaks against an increase in the density of states $N(E_{f})$ as being
responsible for the intrinsic $T_{c}$ enhancement under pressure. \ In fact,
early studies by us of the spin susceptibility of La$_{1.85}$Sr$_{0.15}%
$CuO$_{4}$ \cite{allgeier10}and Y-123 \cite{allgeier11} could detect no change
whatsoever in $N(E_{f})$ under pressure. \ From Eq. (1) it thus follows that
the increase in $T_{c}$ under pressure must come from an increase in the
coupling strength $\mathcal{V}$ itself.

To go further in the analysis it is necessary to ascertain what structural
feature in the highly anisotropic oxides is responsible for the increase in
$\mathcal{V}$ under pressure. \ Uniaxial pressure experiments provide
information on the changes occurring in $T_{c}$ for uniaxial stress applied
both parallel and perpendicular to the CuO$_{2}$ planes. \ One of the most
important experiments in this regard was carried out in 1997 by Meingast
\textit{et al.} \cite{meingast10} on Ca-substituted Y-123. \ These studies
revealed that the compression along the $c$ axis (perpendicular to the
CuO$_{2}$ planes) has no effect on $T_{c},$ whereas compression in the
CuO$_{2}$ planes caused the ubiquitous increase in $T_{c}.$ \ A wealth of
further uniaxial pressure experiments support this conclusion. \ This prompted
Wijngaarden \textit{et al.} \cite{wijn1} to state in 1999 that ``Hence, there
is quite some evidence that $\Delta c$ mainly influences doping, while $\Delta
a$ mainly influences the intrinsic $T_{c}$ ''. \ If we now use the available
anisotropy compressibility data to convert the volume dependence $T_{c}\propto
V^{-1.2}$ into a dependence on the in-plane lattice parameter $a $, we obtain
\begin{equation}
T_{c}\propto a^{-4.5},
\end{equation}
so that $T_{c}$ is approximately proportional to the inverse square of the
area of the CuO$_{2}$ planes. \ This is perhaps the single most important
fundamental result of all high pressure experiments on the cuprate oxides.
\ The message is: \ to further enhance $T_{c}$, try to find structures which
are capable of \textit{compressing} the CuO$_{2}$ planes without adding
defects or increasing the buckling of these planes. \ Obviously this may not
be easy to accomplish since these planes are quite stiff and it would be
anticipated that structural attempts to compress the planes could easily cause
them to buckle.

The next question is whether there is any correlation between the value of
$T_{c}$ at ambient pressure and the value of the in-plane lattice parameter
$a$. \ In Fig. 4 it is seen that there is no such correlation. \ The
single-plane material with the highest value of $T_{c}$ (98 K) is Hg-1201
which has the largest value of $a.$ \ In addition, the compound in Fig. 4 with
the lowest value of $T_{c}$ is La$_{1.85}$Sr$_{0.15}$CuO$_{4}$ which has the
smallest value of $a$! \ We note that applying 4 GPa pressure decreases $a$ by
about 1\% which would generate an increase in $T_{c}$ of about 6 K. \ Within
that range of $a$ in Fig. 4 the variation of $T_{c}$ for the compounds listed
is much greater. \ This simply means that the compounds presently available
have not yet played the high-pressure card, i.e. they haven't made use of the
fact that $T_{c}$ will increase appreciably if $a$ is reduced. \ The system
that managed to reduce $a$ the most is La$_{1.85}$Sr$_{0.15}$CuO$_{4}$;
perhaps the reason that it's $T_{c}$ value is the lowest of all systems in
Fig. 4 arises from the strong structural distortions leading to considerable
plane buckling.%

\begin{center}
\includegraphics[
natheight=6.531100in,
natwidth=9.562200in,
height=3.4177in,
width=4.9943in
]%
{figures/Lattice Parameter Graphic.png}%
\end{center}

\noindent{\small Fig. 4. \ Average lattice parameter in the CuO}$_{2}%
${\small \ plane for representative superconducting oxide systems at ambient
pressure.\vspace{0.6cm}}

To summarize, there are at least three principal ways that high external or
lattice (chemical) pressures can be used to help maximize the value of $T_{c}
$ in the superconducting oxides:\vspace{0.3cm}

\noindent1. \ Use pressure to reach optimal doping in system with few defects
in or near the CuO$_{2}$ planes. \ This may be the best (only) way to reach
optimal doping in systems with more than three CuO$_{2}$ planes.\vspace{0.2cm}

\noindent2. \ Use pressure to help flatten the CuO$_{2}$ planes.\vspace{0.2cm}

\noindent3. \ Apply pressure to reduce the area of the CuO$_{2}$ planes,
keeping them flat.\vspace{0.2cm}

\noindent4. \ Use the result $T_{c}\propto a^{-4.5}$ to help identify the
correct theoretical model and then apply this model to help further optimize
the value of $T_{c}.$ \ This will be the subject of a future paper.\vspace{0.8cm}

\noindent{\Large ACKNOWLEDGEMENTS \ \vspace{0.4cm}}

The author would like the following present and past students in his group at
Washington University for their excellent high-pressure work on
superconducting materials: \ S. Deemyad, J. Diederichs, J.J. Hamlin, A.K.
Klehe, S. Klotz, C.W. Looney, S. Sadewasser, and T. Tomita. \ Work at
Washington University supported by NSF grant DMR-0101809.\vspace{1cm}

\noindent{\Large REFERENCES\vspace{0.4cm}}

\end{document}